# Conditional Guided Generative Diffusion for Particle Accelerator Beam Diagnostics

Alexander Scheinker[1,*]

[1]Applied Electrodynamics Group, Los Alamos National Laboratory, Los Alamos, New Mexico, 87545, USA
[*]ascheink@lanl.gov

## ABSTRACT

Advanced accelerator-based light sources such as free electron lasers (FEL) accelerate highly relativistic electron beams to generate incredibly short (10s of femtoseconds) coherent flashes of light for dynamic imaging, whose brightness exceeds that of traditional synchrotron-based light sources by orders of magnitude. FEL operation requires precise control of the shape and energy of the extremely short electron bunches whose characteristics directly translate into the properties of the produced light. Control of short intense beams is difficult due to beam characteristics drifting with time and complex collective effects such as space charge and coherent synchrotron radiation. Detailed diagnostics of beam properties are therefore essential for precise beam control. Such measurements typically rely on a destructive approach based on a combination of a transverse deflecting resonant cavity followed by a dipole magnet in order to measure a beam's 2D time vs energy longitudinal phase-space distribution. In this paper, we develop a non-invasive virtual diagnostic of an electron beam's longitudinal phase space at megapixel resolution (1024 × 1024) based on a generative conditional diffusion model. We demonstrate the model's generative ability on experimental data from the European X-ray FEL.

## Introduction

Particle accelerators provide intense high energy charged particle beams for a wide range of scientific studies at otherwise inaccessible length, time, and energy scales. Free electron lasers (FEL) generate bright flashes of coherent light at femtosecond time scales which is incredibly useful for structural biology[1]. FELs have been utilized for a wide range of biological studies including protein crystallography[2–4], with a recent demonstration of single protein-based diffraction from a 14 nm diameter sample[5]. Two-color experiments with polarization control has enabled the use of FELs as tools for chiral recognition during photolysis[6]. FELS have been used to image viruses[7], to study the structure and dynamics of macromoleculse[8], and FELs have been used to study matter in extreme conditions[9]. Utilizing FELs as femtosecond light sources has also enabled time-resolved site-specific investigations for understanding and benchmarking ultrafast photochemistry[10]. The data utilized in this work was collected at the European X-ray FEL (EuXFEL)[11]. The EuXFEL is one of the most advanced FEL facilities in the world, capable of accelerating up to 5000 electron bunches per second up to energies of 17.5 GeV, with the FEL undulator producing hard X-rays at up to 14 keV with pulse energies of up to 2.0 mJ. The EuXFEL has been utilized for a wide range of scintific studies. Recent work at the EuXFEL includes the study of ribosome molecules[12], for developing crystal-based photon energy calibration techniques for FELs[13], for the development of advanced single-particle X-ray diffractive imaging techniques[14], for laser-driven dynamic compression experiments for fast formation of nanodiamonds[15], for studies of ultrafast demagnetization induced by X-ray photons[16], for the development of novel single X-ray pulse-based 3D atomic structure reconstructions[17], and for ultrahigh resolution X-ray Thomson scattering[18].

In FELs, photocathode properties are crucial as they define the initial conditions of the electron beams which are then accelerated and used to produce the FEL light. The improvement and development of advanced FEL photocathodes is a lively area of research including a wide range of studies on photocathode technology[19–23]. Another area of intense FEL research is the development of non-destructive characterization methods for the FEL light pulses themselves. This is increibly challenging s the pulses can be only 10s of femtoseconds in duration, but their characterization is crucial to fully understand the FEL-based imaging process and to verify the properties of the produced light. Towards these efforts, recently AI methods have been developed for online characterization of ultrashort X-ray free-electron laser pulses themselves[24].

After the electrons are produced at the photocathode and before they pass through the undulator to create intense pulses of light, the dynamics of intense electron bunches are influenced by complex collective effects such as wakefields, space charge, and coherent synchrotron radiation, making it difficult to control and tune beam properties using model-based approaches. Precisely shaping the 2D energy vs time longitudinal phase space distribution of the electron beam relies on an ability to measure that distribution in detail. The state of the art method for such measurements utilizes a x-band transverse deflecting cavity (XTCAV) to measure the beam. The XTCAV streaks the electron bunch, translating longitudinal position to transverse position. The rotated bunch is passed through a vertical dipole causing an energy-dependent curvature of the electron trajectory, providing a measurement of both longitudinal bunch current profile and energy distribution[25]. Several examples of such measurements taken at the European X-ray FEL (EuXFEL) are shown in Figure 1. The images shown are 1024× 1024 pixels with a time resolution of 1.5 fs/pixel (because the electrons are traveling at near light speed that translates into 4.5 $\mu$m/pixel), the energy resolution is 20 keV/pixel, the bunch charge is 0.25 nC, and the bunch energy is 150 MeV.

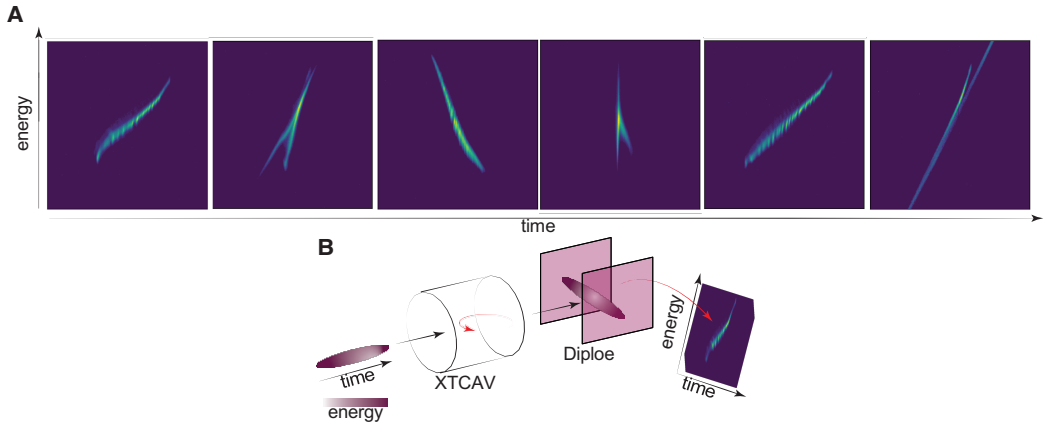

**Figure 1.** **A**: Examples of longitudinal phase space measurements of electron beam at the EuXFEL for various accelerator components settings. **B**: An overview of the destructive measurement process with the beam first rotated by a transverse deflecting RF cavity followed by energy-based dispersion of the beam with a dipole magnet.

The main limitation of a XTCAV-based measurement is that it is an invasive procedure which destroys the beam that is being measured and therefore that same beam cannot be accelerated further for experimental applications. Furthermore, in many facilities (such as the EuXFEL) choosing whether to send the beam off to a diagnostic section or to allow the beam to continue accelerating is a lengthy tuning procedure. It is not possible to simply switch back and forth at the push of a button. Therefore the initial accelerator section essentially runs at two different modes. First, the XTCAV diagnostics section is used for initial tuning, then the beam is only accelerated downstream without further use of the XTCAV.

It would be incredibly valuable to measure the detailed time vs energy distribution of the electron beam near the beginning of the accelerator at all times non-invasively, both to provide a detailed understanding of the electron beam's characteristics and also to use that information as input to online physics models which could then realistically estimate the beam's dynamics through subsequent accelerator sections. These days, various machine learning (ML)-based methods for particle accelerators, including for use as virtual diagnostics have been studied for many accelerator applications. For example, neural networks are being used for uncertainty aware anomaly detection to predict errant beam pulses[26].

In terms of virtual diagnostics, encoder-decoder neural networks were developed for the EuXFEL for generating longitudinal ($z, E$) phase-space images of the electron beam[27]. Neural networks have been used to develop virtual diagnostics for 4D tomographic phase space reconstructions[28]. Neural network-based methods have been developed for predicting the transverse emittance of space charge dominated beams[29]. Adaptive neural networks using advanced feedback control algorithms[30] for adaptive latent space tuning of autoencoders have been developed to provide virtual 6D diagnostics of charged particle beams[31], and these adaptive ML methods have been shown to increase the robustness of generative predictions far beyond the span of the training data, for tracking unknown time-varying beams[32]. Adaptive ML methods have also been designed for inverse problems that map downstream beam measurements back to the initial beam distribution[33]. Recently, very interesting methods have also been studied for phase space reconstructions based on normalizing flows[34].

In this paper, the first diffusion based approach to non-invasive high resolution beam diagnostics is introduced. The diffusion-based model is developed for imaging the time vs energy longitudinal phase space distribution of a charged particle beam, and demonstrated on the European XFEL for accurately predicting the distributions of a diverse set of bunch profiles over a wide range of accelerator settings. Although this method is focused on a particle accelerator application, it is a very general approach, which can be used for any complex dynamic system for which it would be beneficial to replace invasive or destructive diagnostics of the system's state with virtual diagnostics which must rely on low-dimensional non-invasive measurements or parameter set points, as shown in Figure 2.

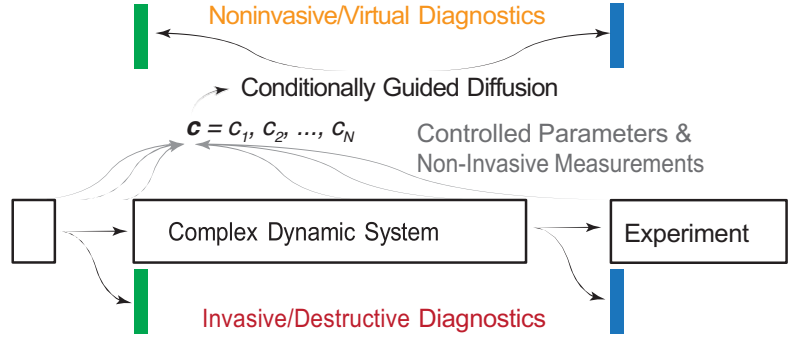

**Figure 2.** A general high-level overview of using conditional diffusion as a generative model that provides a non-destructive detailed view of the state of a complex system based on any available non-invasive diagnostics.

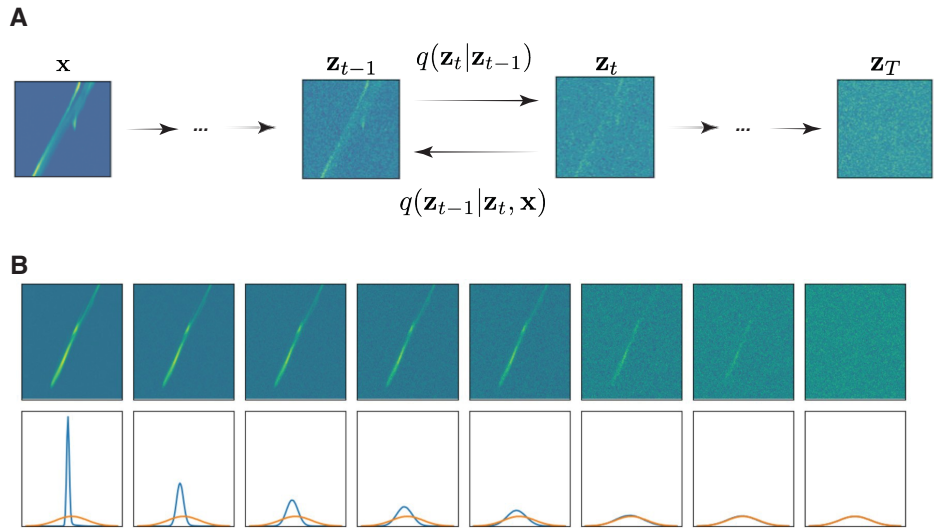

**Figure 3.** **A**: The forward diffusion process of a sample **x** is shown as it transforms to latent variables $z_t$ which gradually approach a Gaussian distribution. **B**: A $T$ = 1000 step diffusion process is shown at each 125 steps along with a histogram of the image pixel values at each step (blue) next to a mean zero unit variance Gaussian (yellow).

## Methods

Generative models based on diffusion utilize a gradual denoising approach inspired by statistical thermodynamics for modeling complex distributions[35]. This approach was then further developed for the generation of high resolution images[36–39]. Diffusion-based generative models have now become the state-of-the-art for generating high resolution images, especially when the images have a wide variety. The generative ability of diffusion-based models has made them powerful tools for a wide range of scientific applications[40], such as conditional generation of hypothetical new famlies of superconductotrs[41], for brain imaging[42], for various bioengineering applications[43], for protein structure generation[44].

The diffusion-based method used in this work is based on a modified version of the approach described in[36]. In our approach we add an additional conditional input vector along with the time embedding to perform guided diffusion which maps specific accelerator conditions to beam images. For completeness, we briefly outline the generative diffusion theory and refer readers to[36] for details and proofs.

The generative diffusion process works by adding noise to an image **x** over a large number of steps $T$ until the pixel values of the image closely resemble a mean zero unit variance Gaussian distribution as shown in the left to right flow of Figure 3.

For an image **x**, the first step of the diffusion process is to create a noise-corrupted image $z_1$ as defined by

$$\mathbf{z_1} = \sqrt{1-\beta_1}\mathbf{x} + \sqrt{\beta_1}\epsilon_1, \quad \epsilon_1 \sim \mathcal{N}(\epsilon_1|0,I) \quad (1)$$

with subsequent diffusion steps iteratively defined as

$$z_t = \sqrt{1 - \beta_t} z_{t-1} + \sqrt{\beta_t} \epsilon_t, \quad \epsilon_t \sim \mathcal{N}(\epsilon_t|0, I), \quad t \in \{2, ..., T\} \tag{2}$$

The noise schedule βt $\beta_t$ [0, 1] with $\beta_1 < \beta_2 < ... < \beta_T$ prescribes the variance for the additive unit variance Guassian noise $\varepsilon_t$ at each step $t$ which defines how quickly images are converted to pure noise. This diffusion sequence forms a Markov chain with conditional distributions of the form

$$q(z_t|z_{t-1}) = \mathcal{N}(z_t|\sqrt{1 - \beta_t} z_{t-1}, \beta_t I), \tag{3}$$

which is convenient for sampling random diffusion steps $t$ without having to re-run the entire chain as $z_t$ can be rewritten as

$$z_t = \sqrt{\alpha_t} x + \sqrt{1 - \alpha_t} \epsilon_t \tag{4}$$

where $\sqrt{1 - \alpha_t} \epsilon_t$ represents the total noise added to the original image and $\alpha_t$ is the product.

$$\alpha_t = \prod_{\tau=1}^{t}(1 - \beta_\tau) \tag{5}$$

Equation (4) implies that

$$q(z_t|x) = \mathcal{N}(z_t|\sqrt{\alpha_t} x, (1 - \alpha_t)I), \tag{6}$$

and therefore, since $(1 - \beta_t) < 1$, as $T \to \infty$ the terms $\alpha_t$ and $1 - \alpha_t$ approach 0 and 1, respectively, and

$$\lim_{T \to \infty} q(\mathbf{z}_T | \mathbf{x}) = \mathcal{N}(\mathbf{z}_T | \mathbf{0}, \mathbf{I}), \tag{7}$$

which means that any image is converted to a signal indistinguishable from mean 0 unit variance Gaussian noise. In practice values such at $T = 1000$ are a good choice, which is also the number of diffusion steps used in this work. The noise schedule is chosen with endpoints as in[36] with $\beta_t$ increasing linearly from $10^{-4}$ to 0.02 over $T = 1000$ steps. While nonlinear noise schedules were proposed in[38], the authors pointed out that they were mostly beneficial for lower resolution images, and that was confirmed in this work as a cosine noise schedule performed similarly to the linear one.

In order to generate images, the model must learn to run backwards, undoing the diffusion process. For a given image **x**, Baye's rule and the Gaussian change of variables identity allow us to write the conditional probability $q(\mathbf{z}_{t-1} | \mathbf{z}_t, \mathbf{x})$ which describes one reverse step of the diffusion process, as:

$$q(\mathbf{z}_{t-1}|\mathbf{z}_t, \mathbf{x}) = \mathcal{N}\left(\mathbf{z}_{t-1} \middle| \frac{1 - \alpha_{t-1}}{1 - \alpha_t}\sqrt{1 - \beta_t}\mathbf{z}_t + \frac{\sqrt{\alpha_{t-1}} \beta_t}{1 - \alpha_t}\mathbf{x}, \frac{\beta_t(1 - \alpha_{t-1})}{1 - \alpha_t}\mathbf{I}\right). \tag{8}$$

By rewriting Equation (4) as

$$x = \frac{1}{\sqrt{1 - \beta_t}} z_t - \frac{\sqrt{\beta_t}}{\sqrt{1 - \beta_t}} \epsilon_t, \tag{9}$$

and plugging that into the mean value of the prediction in Equation 8, the model can be effectively used to predict how to remove noise between iterative steps to restore the original image according to

$$z_{t-1} = \frac{1}{\sqrt{1 - \beta_t}}\left[z_t - \frac{\beta_t}{\sqrt{1 - \alpha_t}} D(z_t, t, c)\right] + \sqrt{\beta_t} \epsilon, \quad \epsilon \sim \mathcal{N}(\epsilon|0, 1). \tag{10}$$

At the final generation step, the image is created according to

$$x = \frac{1}{\sqrt{1 - \beta_1}}\left[z_1 - \frac{\beta_1}{\sqrt{1 - \alpha_1}} D(z_1, t, c)\right]. \tag{11}$$

One example of such diffusion-based image generation is shown in Figure 4.

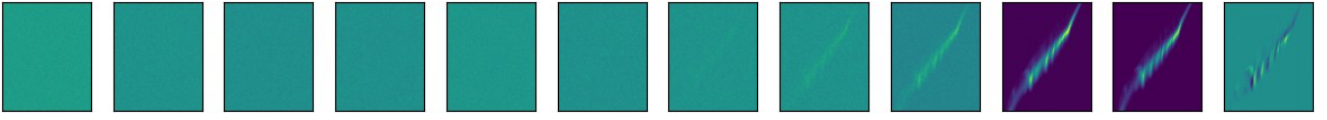

**Figure 4.** From left to right the first ten images show diffusion steps 100, 200, ..., 1000 followed by the true target image and finally the difference between the two.

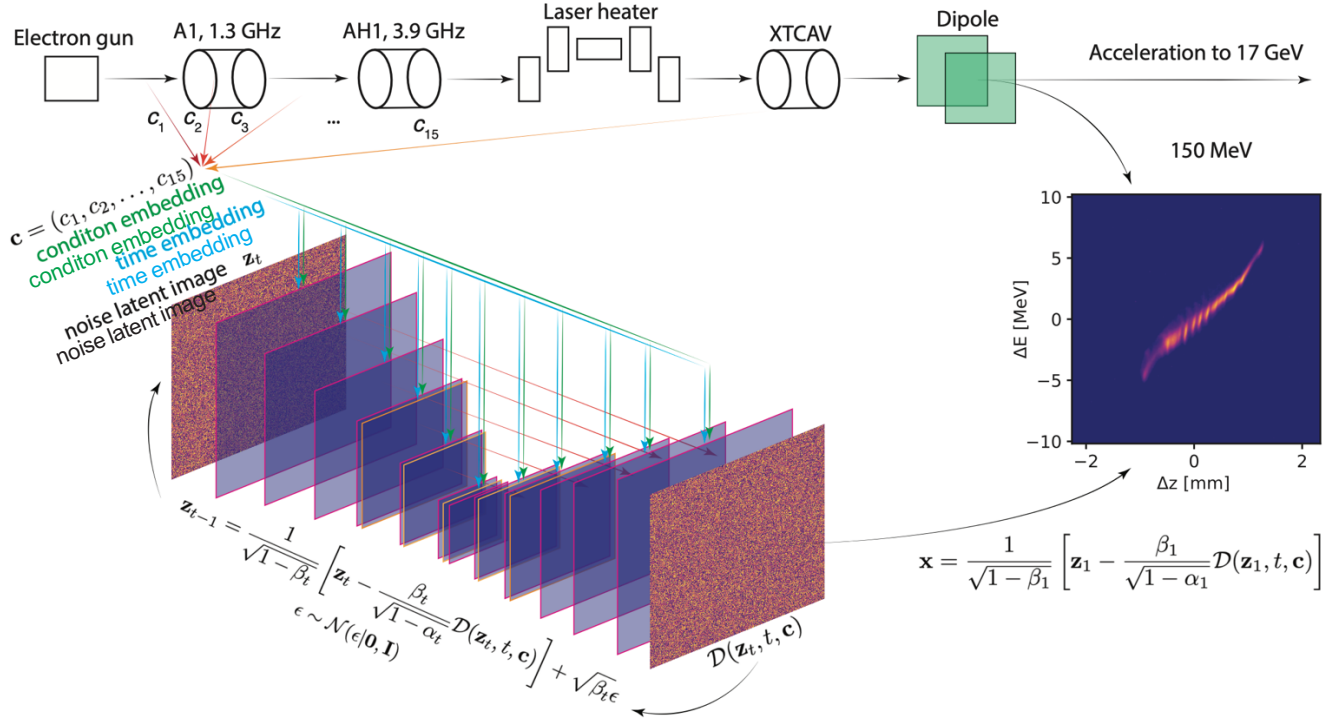

**Figure 5.** Conditional diffusion setup for generating XTCAV images of the beam's $(z, E)$ longitudinal phase space projection.

## Results

This work demonstrates that a conditionally guided generative diffusion process can be used to accurately generate unseen test data to give a non-invasive virtual high resolution view of the electron beam's longitudinal $(z, E)$ phase at the EuXFEL. In this approach, the conditional input vector is $\mathbf{c} \in \mathbb{R}^{15}$ utilizing 5 accelerator parameter set points and 10 non-invasive beam-based measurements, as shown in Figure 5. The first 5 parameters of $\mathbf{c}$ are settings of three energy chirps (energy vs time correlations designed for bunch compression) imposed by radio frequency (RF) resonant cavity fields onto the electron beam as it passes from the electron gun through the A1 and AH1 sections of the accelerator. The curvature of this chirp is also controlled and so is the third derivative by using 3.9 GHz RF which is at the third harmonic of the EuXFEL's overall 1.3 GHz RF system. The remaining 10 parameters are RMS $X$ and $Y$ beam centroid values measured at 5 locations between the injector and the XTCAV. The data used for this work was gathered at the EuXFEL by varying the first 5 parameters above randomly within a wide range. From 11000 data points, 10000 were used for training and 1000 for testing.

This generative conditionally guided diffusion approach results in an ability to create incredibly high resolution megapixel beam images for a very wide range of beams, which is exactly the application for which diffusion-based models are state-of-the-art. Figure 7 shows three detailed examples of very different very complex beams generated by the conditional diffusion process next to the true images.

To quantify the reconstruction accuracy the absolute percent error was calculated between generated images $\hat{Y}$ and their true measurements $Y$ according to

$$E = 100 \times \sum_i \sum_j |Y - \hat{Y}| / \sum_i \sum_j |Y|, \tag{12}$$

where $i, j \in \{1, 1024\}$ are the pixel locations within the images. Figure 6 quantitatively shows the generative error as defined in

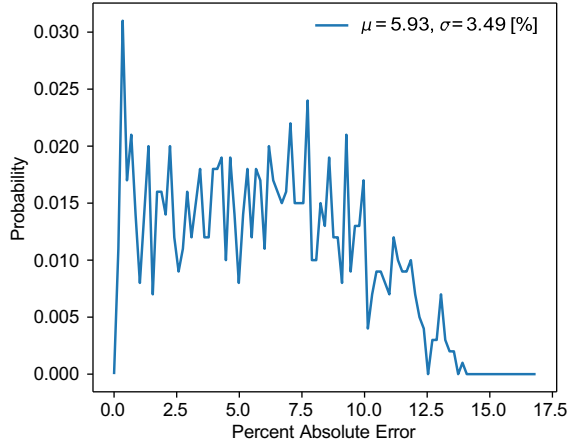

**Figure 6.** Error statistics for 1000 test images. The overall mean absolute percent error was 5.93 % with a standard deviation of 3.49 %.

Equation [12](#) for 1000 unseen test objects. For a visualization of how these error levels correspond to image quality, Figure [8](#) shows 5 examples from the test data set, of increasing error. It can be seen that up to ∼ 10 % error the predictions are very accurate.

These results demonstrate that the conditionally guided diffusion model can serve as a highly accurate virtual diagnostic of the electron beam's longitudinal phase space for a wide range of accelerator settings, without having to intercept and destroy the beam to measure it.

### Moving Along the Image Manifold by Latent Space Interpolation

Once the diffusion model has been trained to generate a electron beam images associated with a wide range of accelerator settings, it is possible to smoothly move between various accelerator setups while generating realistic electron bunch distributions. Given two different accelerator setups ($c_0$, $c_1$) and their associated electron beam images ($x_0$, $x_1$), it would be useful to understand how the beam behaves at an intermediate state between these two. One simple naive way to attempt to approximate this is a linear interpolation between the two images of the form

$$\mathbf{x}_t = (1-t)\mathbf{x}_0 + t\mathbf{x}_1, \quad t \in [0, 1], \quad t : 0 \to 1 \quad \Longrightarrow \quad \mathbf{x}_t : \mathbf{x}_0 \to \mathbf{x}_1. \tag{13}$$

This method results in a non-physical weighted superposition of the two images, as shown on the left side of Figure [9](#).

The trained diffusion model allows us to move between accelerator settings in a more physical way, as the network utilizes all of the training data in order to interpolate in a physically consistent way. In this approach, for the two images ($x_0$, $x_1$) we first perform a conditional generation of the two images based on two random noise images ($n_0$, $n_1$) and on their two conditioning vectors which correspond to accelerator settings ($c_0$, $c_1$). Together we treat the noise-vector pairs as latent variables ($z_0$, $z_1$). We now perform the same linear interpolation as above, but in the latent space directly according to

$$\mathbf{z}_t = (1-t)\mathbf{z}_0 + t\mathbf{z}_1, \quad t \in [0, 1], \quad t : 0 \to 1 \quad \Longrightarrow \quad \mathbf{z}_t : \mathbf{z}_0 \to \mathbf{z}_1. \tag{14}$$

At each time step along the latent path $\mathbf{z}_t$ we can then perform conditional generation via the learned diffusion process in order to generate an image $\mathbf{x}_t$ which is no longer a simple superposition of the two images, but rather is a true intermediate state that is moving along the learned image manifold

$$\mathbf{x}_t = D(\mathbf{z}_t) = D\left((1-t)\mathbf{z}_0 + t\mathbf{z}_1\right), \tag{15}$$

as shown on the right side of Figure [9](#).

### Discussion

This paper has demonstrated that conditionally guided generative diffusion models can be utilized as high resolution virtual diagnostics for charged particle beams. It was shown that the models can make accurate predictions for unseen test data within a very diverse set of measurements, and that the trained models can be used to smoothly traverse the learned latent embedding in order to interpolate between various accelerator settings in a physical way. Although this work was focused on high resolution 2D ($z$, $E$) longitudinal phase space predictions, because those are some of the most important measurements for FEL operations, this same approach can be used to model any of the beam's projections, including all of the 15 unique 2D projections of a beam's 6D phase space as was already done with autoencoder-based generative models[31,32]. The benefit of using diffusion for this approach should be in generating higher resolution images.

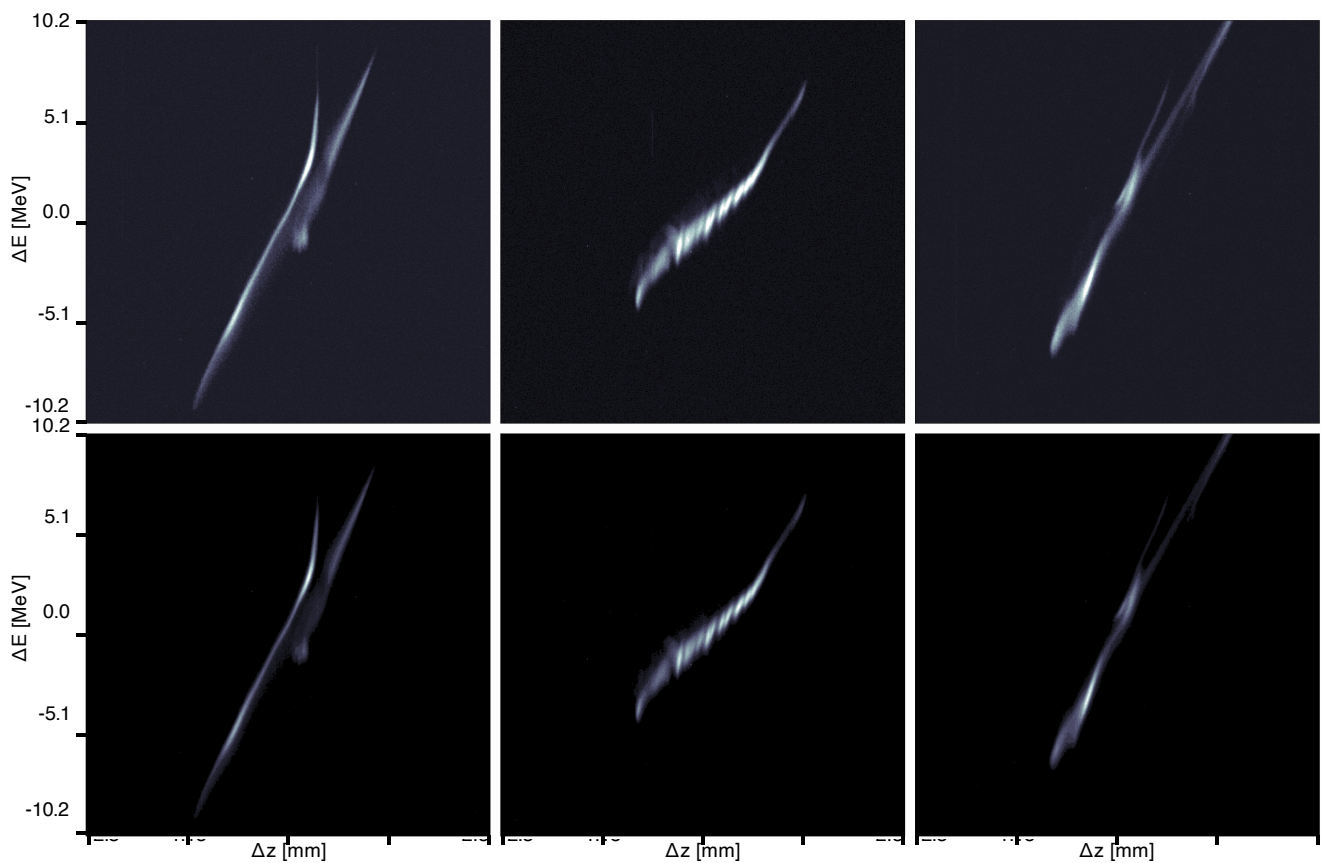

**Figure 7.** The top row shows a detailed view of 3 conditionally generated electron beam images at 1024 ×1024 pixel resolution, based on a 1000-step diffusion process. The bottom row shows the target electron beam images.

Beyond particle accelerators, such a conditionally guided diffusion process could be very useful for any complex dynamic system in which it would be beneficial to replace a destructive / invasive diagnostic of the system state with a virtual non-invasive virtual diagnostic.

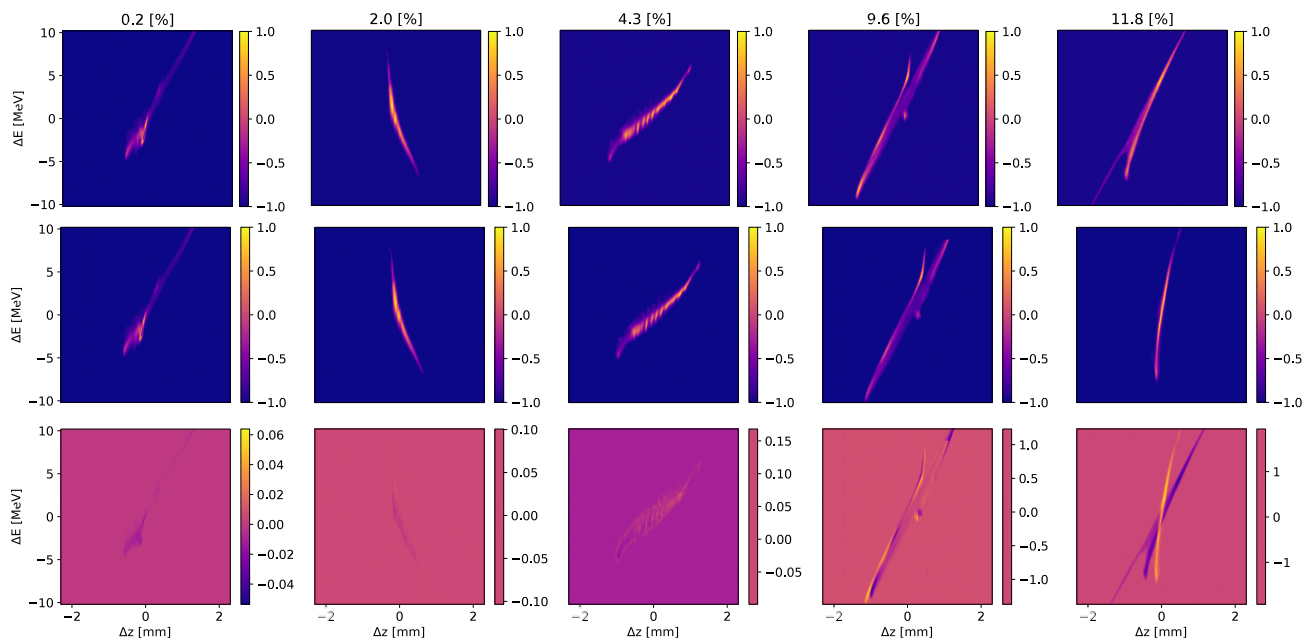

**Figure 8.** Detailed view of 5 reconstructed test images of increasing percent absolute error from left to right.

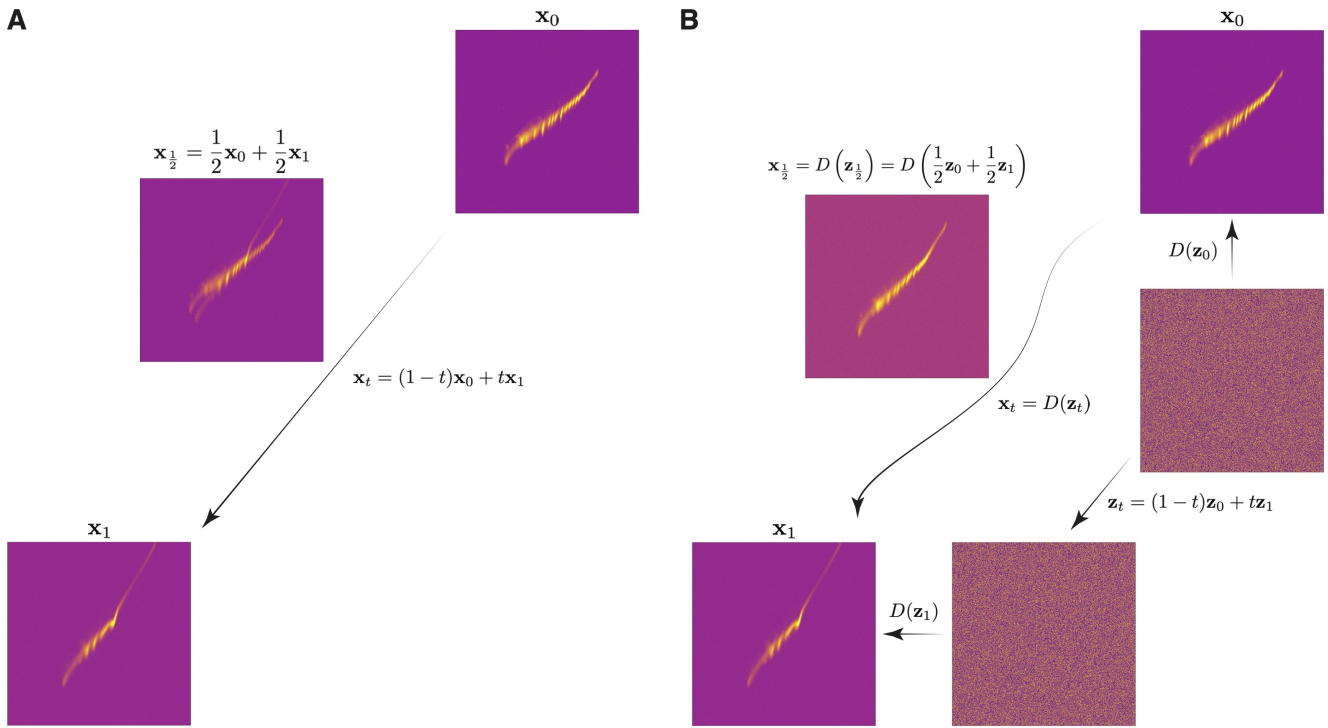

**Figure 9.** Left: Performing a simple linear interpolation between two images of the electron beam ($\mathbf{x}_0$, $\mathbf{x}_1$) at two different accelerator settings results. Right: Performing linear interpolation in the latent space.

## Acknowledgements

This work was supported by the U.S. Department of Energy (DOE), Office of Science, Office of High Energy Physics and the Los Alamos National Laboratory LDRD Program Directed Research (DR) project 20220074DR. This research used resources provided by the Los Alamos National Laboratory Institutional Computing Program, which is supported by the U.S. Department of Energy National Nuclear Security Administration under Contract No. 89233218CNA000001.

## Competing interests

The author declares no competing interests.